\documentclass{neuthist18}

\usepackage{wrapfig}

\usepackage{enumitem}
\setenumerate{noitemsep,topsep=0pt,parsep=0pt,partopsep=0pt}

\usepackage{amsmath}

\usepackage{array}

\usepackage[symbol]{footmisc}

\def\Journal#1#2#3#4{{#1} {\bf #2}, #3 (#4)}

\def\PLB{{\em Phys. Lett.}  B}
\def\PRL{\em Phys. Rev. Lett.}
\def\PRD{{\em Phys. Rev.} D}

\def\EPJC{{\em Eur. Phys. J.} C}
\def\NPBPS{\em Nucl. Phys. B (Poc. Suppl.)}

\def\nova {NO$\nu$A}
\def\etal{{\it et al.}}
\def\numu{\ensuremath{\nu_\mu}}
\def\nue{\ensuremath{\nu_e}}
\def\nutau{\ensuremath{\nu_\tau}}

\def\nuebar{\ensuremath{\bar{\nu}_e}}
\def\sinsqtwo#1{\ensuremath{\sin^{2}2\theta_{ #1 }}}
\def\dynamic{\ensuremath{\sin^2(\Delta m^2 L/4E)}}
\def\dynamicx#1{\ensuremath{\sin^2(\Delta m^2_{ #1 } L/4E)}}
\def\dmsq#1{\ensuremath{\Delta m^2_{ #1 }}}
\def\sinsq#1{\ensuremath{\sin^2\theta_{ #1 }}}

\newcommand{\Photo}{}

\begin{document}
\vspace*{3cm}

\title{History of Long-Baseline Accelerator Neutrino Experiments\footnote[1]{Invited talk at the Conference on the History of the Neutrino, Paris, 5-7 September 2018} }

\author{ Gary J. Feldman }

\address{Department of Physics, Harvard University, 17 Oxford Street,\\
Cambridge, MA 02138, United States}

\maketitle\abstracts{
I will discuss the  six previous and present long-baseline neutrino experiments: two first-generation general experiments, K2K and MINOS, two specialized experiments, OPERA and ICARUS, and two second-generation general experiments, T2K and \nova.     The motivations for and goals of each experiment, the reasons for the choices that each experiment made, and the outcomes will be discussed.}

\section{Introduction}

My assignment in this conference is to discuss the history of the six past and present long-baseline neutrino experiments.  Both Japan and the United States  have hosted first- and second-generation general experiments,   K2K and T2K in Japan and MINOS and \nova\ in the United States.  Europe hosted two more-specialized experiments, OPERA and ICARUS.  Since the only possible reason to locate a detector hundreds of kilometers from the neutrino beam target is to study neutrino oscillations, the discussion will be limited to that topic, although each of these experiments investigated other topics.  Also due to the time limitation, with one exception, I will not discuss sterile neutrino searches.  These experiment have not found any evidence for sterile neutrinos to date.\cite{sterileone,steriletwo,sterilethree,sterilefour,sterilefive,sterilesix,sterileseven}

\begin{wrapfigure}{r}{0.40\textwidth}
\vspace{-15pt}
\centering
\includegraphics[width=0.35\textwidth]{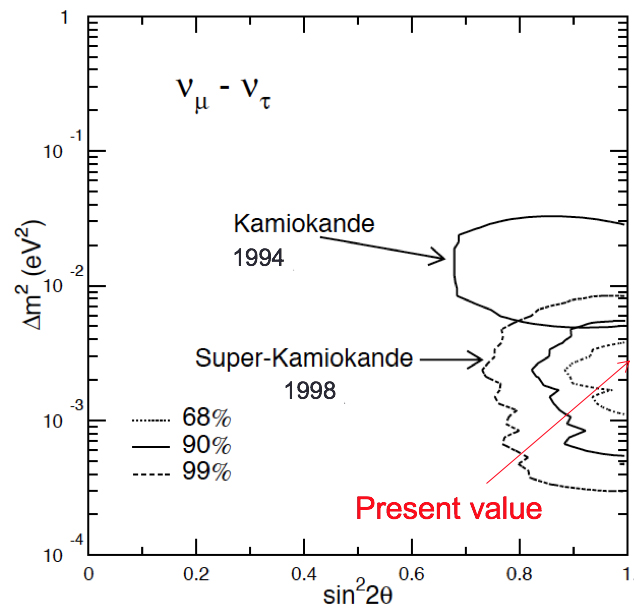}
\caption{\label{fig:kamiokande}$\Delta m^2$ versus $\sin^{2}2\theta$ contours for the
Kamiokande\ and Super-Kamiokande experiments.}
\vspace{-20pt}
\end{wrapfigure}

The Japanese and American hosted experiments have used the comparison between a near and far detector to measure the effects of oscillations.  This is an essential method of reducing systematic uncertainties, since uncertainties due flux, cross sections, and efficiencies will mostly cancel.  The American experiments used functionally equivalent near detectors and the Japanese experiments used detectors that were functionally equivalent, fine-grained, or both.
The use of fine-grained near detectors for water Cherenkov far detectors are quite useful since the water Cherenkov detectors are not sensitive to particles with velocities smaller than the Cherenkov threshold.

\section{First-Generation Experiments}

\subsection{Motivation}

The proposals for the first generation of long-baseline experiments were submitted in 1994, motivated by the atmospheric results from deep underground experiments whose {\it raisons d'\^{e}tre} were the searches for proton decay.  In 1994, the Kamiokande experiment produced clear evidence for neutrino oscillations as shown as a contour in the $\Delta m^2$ versus $\sin^{2}2\theta$ plot shown in Fig.~\ref{fig:kamiokande}.\cite{kamiokande}   This figure also shows the higher statistics Super-Kamiokande data from 1998 \cite{superk} as well as the presently accepted value of \dmsq{23}.  Note that the Kamiokande central $\Delta m^2$ value was roughly an order of magnitude higher than the present value, which, as we will see, had an effect on the planning of the first generation long-baseline experiments.  Even with the Super-Kamiokande results of 1998, there was a widespread uncertainty in the value of \dmsq{23}.  

\begin{figure}[htb]
\includegraphics[width=\linewidth]{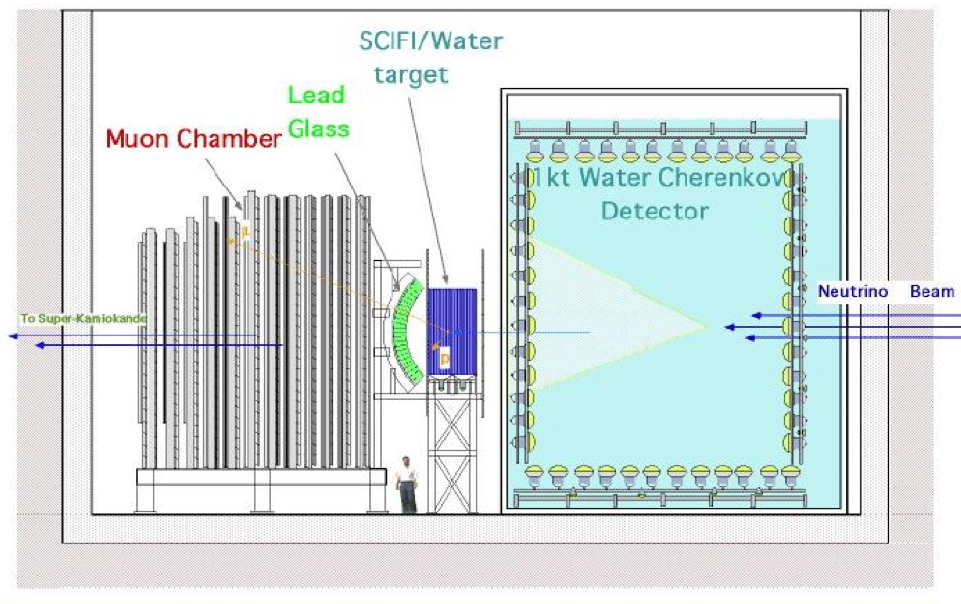}
\caption{\label{fig:k2knd} K2K near detector.}
\end{figure}

\subsection{The K2K Experiment}

The K2K (KEK to Kamioka) experiment used the KEK 12 GeV synchrotron in Tsukuba, Japan to send a  neutrino beam with average energy of 1.4 GeV 250 km to the 50 kt Super-Kamiokande water Cherenkov detector.\cite{k2k}   The K2K near detector, shown in Fig.~\ref{fig:k2knd} was located 300 m downstream from the target and was  composed of both functionally equivalent and fine-grained detectors: a 1-t water Cherenkov detector followed by  a scintillating fiber water sandwich detector, a lead glass array, and  a muon ranger.

In common with all accelerator-produced neutrino beams, the K2K neutrino beam was composed of about 99\% \numu\ and 1\% \nue, the latter from muon and kaon decay. Thus, the attempted oscillation measurements would be \numu CC disappearance and \nue CC appearance.  In a 2-neutrino model, appropriate for this period, these oscillation modes would be determined by
\begin{align}
&P(\numu\rightarrow\numu)= 1-\sinsqtwo{\mu\mu}\dynamic, \\
&P(\numu\rightarrow\nue)= \sinsqtwo{\mu e}\dynamic.
\end{align}

K2K realized that it would not be competitive in the long run with the MINOS detector that was being proposed at the same time.  However, it could be the first accelerator experiment to confirm the Kamiokande result, and, after 1998, resolve the difference in the two different $\Delta m^2$ mesurements.\cite{k2kproposal}

\begin{figure}[htb]
\includegraphics[width=0.95\linewidth]{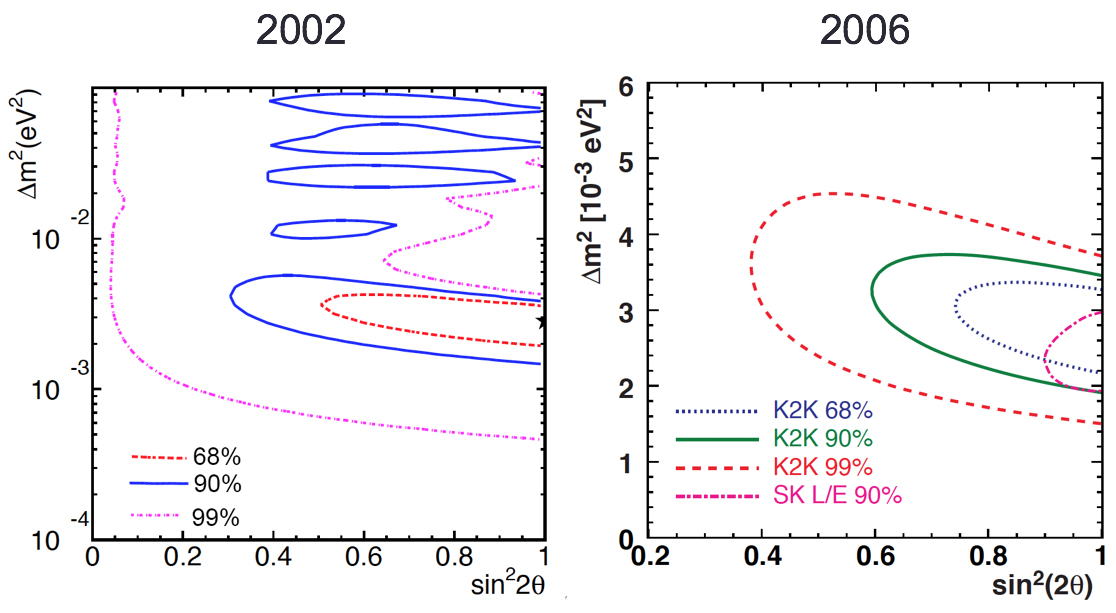}
\caption{\label{fig:k2kresults} Initial and final K2K results on \numu disappearance.}
\end{figure}

K2K ran from June 1999 to November 2004 and accumulated $0.9 \times 10^{20}$ protons on target (POT).\cite{k2kfinal}  First \numu\   disappearance results with a $\Delta m^2-\sin^2(2\theta)$ plot appeared in 2003\,\cite{k2kfirst} and the final results were reported in 2006.\cite{k2kfinal}.  Both of these results are shown in Fig.~\ref{fig:k2kresults}.   Note the good agreement in the final data of the K2K and Super-Kamiokande results.  A search for \nue\ appearance  found one candidate with an expected background of 1.7, setting a 90\% C.L. limit of $\sinsqtwo{\mu e} < 0.13$, about 3 times the current effective value.\cite{k2knue}. 

\subsection{The MINOS Experiment}

The first proposal for a long-baseline neutrino oscillation experiment at Fermilab came in 1991 and was updated in 1993.  The Soudan collaboration proposed a beam to their 1-kt finely segmented iron calorimeter in the Soudan Mine in Northern Minnesota, which was studying atmospheric neutrino events and proton decay.\cite{soudan}  The Fermilab Program Advisory Committee (PAC) felt that a 1-kt detector was too small and called for more ambitious  expressions of interest (EOIs] in 1994.\cite{PACcall}  Three EOIs were received, one from the Soudan group,\cite{sudanprop}  one from US groups working on the MACRO experiment,\cite{MACROprop} and a one-person EOI from Stan Wojcicki.\cite{wojprop}  The PAC asked the three groups to get together and submit a single proposal, which they did in 1995.\cite{MINOSprop}

\begin{wrapfigure}{r}{0.45\textwidth}
\vspace{-15pt}
\centering
\includegraphics[width=0.40\textwidth]{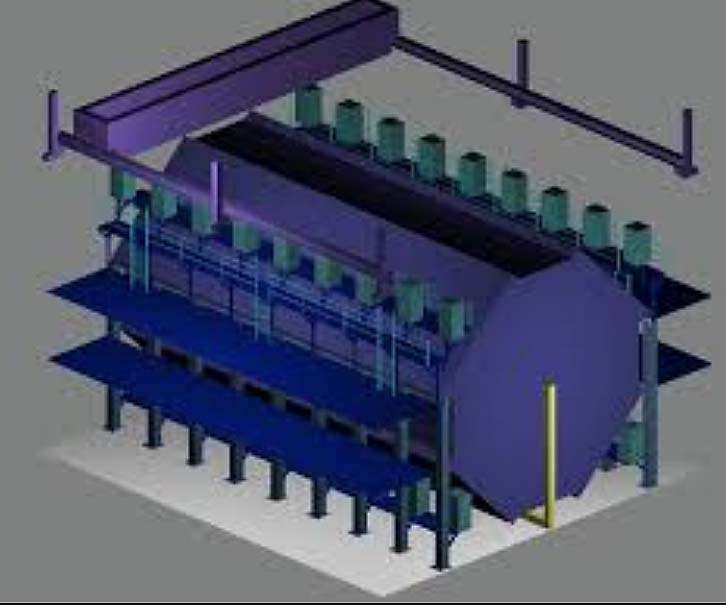}
\caption{\label{fig:minos}MINOS far detector.}
\vspace{-20pt} 
\end{wrapfigure}

The MINOS (Main Injector Neutrino Oscillation Search) proposal called for a 10-kt toroidally magnetized iron-plastic limited streamer tube sandwich detector, with 4-cm thick iron plates and 1-cm tubes.  What was actually built was a 5.4-kt detector with 2.54-cm thick iron plates and 4.1-cm wide plastic scintillator strips.  A drawing of the far detector is shown in Fig.~\ref{fig:minos}.

The far site was chosen to be the Soudan mine, which gave a 735-km baseline.  The rationale for this choice was that the Soudan detector was there and could be used if the construction of the MINOS detector was delayed\,\cite{PACcall}.  Ironically, the MINOS detector was completed two years before the Fermilab NuMI (Neutrinos from the Main Injector) beam was ready.

There was also a 1994 proposal from Brookhaven National Laboratory (E889) for a “long” baseline experiment that had 4 identical 4.5-kt water Cherenkov detectors located at 1, 3, 24, and 68 km from the target, all at an angle of $1.5^\circ$ to the neutrino beam.\cite{Brookhaven}   This off-axis configuration would produce a narrow band beam focused (ideally) on the oscillation maximum, a scheme that would later be adopted by both the T2K and \nova\ experiments.  However, note that this proposal was using the Kamiokande measurement of $\Delta m^2$, which would have focused the beam at an $L/E$ an order of magnitude too high.  The moral of this story is that search lights are useful if you know where to point them.

In 1995, the High Energy Physics Advisory Panel (HEPAP) requested a subpanel chaired by Frank Sciulli to make recommendations on neutrino oscillation experiments.  In September 1995, the subpanel unanimously recommended that MINOS be supported and that Brookhaven E889 not be supported.\cite{Sciulli}   The subpanel gave three reasons for its recommendation:
\begin{enumerate}
  \item MINOS had better sensitivity at low values of $\Delta m^2$.
  \item MINOS had better exploration capabilities.
  \item MINOS would use the same beamline as COSMOS,  an approved short baseline experiments searching for oscillations into heavy {\nutau}'s, which at the time were thought to be dark matter candidates.  Clearly building one beamline would be more efficient than building two.  This last reason turned out to be irrelevant since COSMOS was never funded. 
\end{enumerate}

MINOS took data from February 2006 to April 2012, accumulating $10.7 \times 10^{20}$ POT in neutrino mode and $3.4 \times 10^{20}$ POT in antineutrino mode.  The first publication on \numu\ disappearance was in 2006\,\cite{MINOSfirst} and the final results were reported in 2014.\cite{MINOSfinal}.  The MINOS detector with 1.4-radiation length iron plates was not optimum for measuring \nue\ appearance.  However, MINOS was able to pull out a 1.8 standard deviation result in neutrino mode, 152 candidate events with an expected (neutral-current dominated) background of $128 \pm 5$ (syst.) events.\cite{MIONSnue}  In antineutrino mode, there were 20 candidates with an expected background of 17.5 with an uncertainty due to systematics of 0.8.

Table \ref{tab:compare}, near the end of this talk, will show a quantitative comparison of the four general experiments.

\begin{wrapfigure}{r}{0.40\textwidth}
\centering
\vspace{-20pt} 
\includegraphics[width=0.35\textwidth]{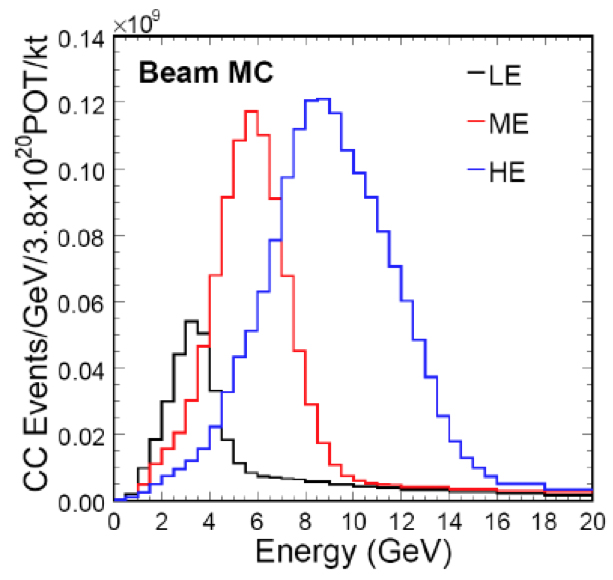}
\caption{\label{fig:NUMI}Fermilab neutrino beam high, medium, and low energy tunes.}
\end{wrapfigure}

\subsection{The MINOS+ Experiment}

The Fermilab NUMI neutrino beam had the ability to  change the beam energy tune, as shown in Fig.~\ref{fig:NUMI}, by moving the distance of the target with respect to the first magnetic horn and optionally the position of the second horn.  The MINOS experiment ran almost entirely in the low energy tune, since that produced the most events in the region of the maximum oscillation.  However, the \nova\ experiment ran with the medium energy tune, since that produced the best sensitivity for \nova's off-axis beam.   Thus in 2013, with the start of \nova\ data taking in the medium energy tune, MINOS reorganized as MINOS+ with the main goals of searching for non-standard interactions and sterile neutrinos.\cite{sterileone,steriletwo}  MINOS+ ended data taking in June, 2016.

\section{The European Program}

\subsection{Motivation}

In 1999, CERN committed itself to CNGS (CERN Neutrinos to Gran Sasso), a program whose main focus was the measurement of \nutau\  appearance.\cite{CERNcouncil}  I have often wondered why Europe would choose to invest in this program.  It was obvious to me at the time, and to many other people,\cite{heuer} that this program would not be competitive with what we would learn better from other measurements in a comparable time period. The argument is simple:
\begin{enumerate}
\item We knew from LEP that there are only 3 active neutrino species.\cite{LEPone,LEPtwo,LEPthree,LEPfour}
\item We knew from Super-Kamiokande that \nue\ appearance was small.\cite{superk}
\item This left only $\tau$ neutrinos and sterile neutrinos.  And MINOS would search for sterile neutrinos with higher precision by searching for neutral current disappearance.\cite{sterileone}
\end{enumerate} 

\noindent The decision to go in this direction was probably due to a combination of the following reasons.

\begin{enumerate}
\item The impractability of building a near detector, since the only possible place was under the Geneva airport.  An official CERN document claims that a near detector is not needed for an appearance measurement.\cite{CERNTDR}  As a generic statement, this is not true.  However, in this case it was true, because the statistical uncertainties would be large.
\item The desire to do something different from Japan and the United States.
\item The interest of Italian and other physicists in pursuing emulsion technology.
\item The pressure from Italy to enhance the Gran Sasso program.  Italian sources were to pay for 68\% of the marginal cost.\cite{CERNcouncil}
\item The conviction that \nutau\ appearance is a crucial measurement.
\end{enumerate}

\subsection{The OPERA Experiment}

\begin{figure}[htb]
\centering
\includegraphics[width=0.75\linewidth]{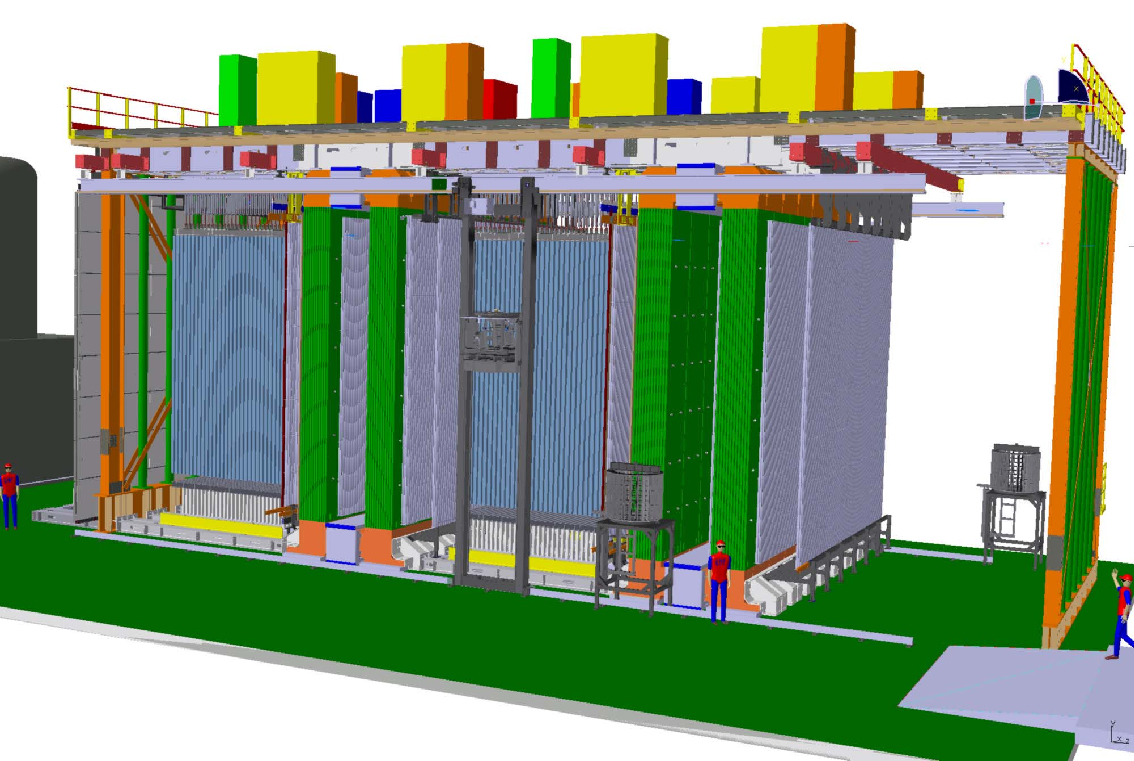}
\caption{\label{fig:opera} OPERA detector.}
\end{figure}

The OPERA (Oscillation Project with Emulsion-tRacking Apparatus) detector was composed of 1.25 kt of sandwiches of 1-mm lead plates and emulsion films to identify \nutau’s by observing kinks from their decays in the emulsion films.  Scintillator trackers identified location of the events in the lead-emulsion ``bricks,'' which could be removed for analysis.\cite{OPERA}  The experimental was located 730 km from the CERN SPS neutrino beam target, ran from 2008 to 2012, and accumulated $1.8 \times 10^{20}$ POT.  A drawing of the detector is shown in Fig.~\ref{fig:opera}.

To optimize \nutau\ appearance, the neutrino beam was tuned to an average neutrino energy of 17 GeV.  This yielded an $L/E$ more than an order of magnitude smaller than the oscillation maximum, resulting in a more than two orders of magnitude suppression from the dynamic term in the oscillation formula, \dynamic.   

The OPERA experiment reported its first \nutau\ event in 2010\,\cite{OPERAfirst} and a total of 5 events in 2015.\cite{OPERAfinal}  Recently, OPERA reported a new analysis with looser selection criteria.\cite{OPERA}  Ten \nutau\ candidates were found with an expected background of $2.0 \pm 0.4$ events and an expected signal of $6.8 \pm 0.8$ events. 

\subsection{The ICARUS Experiment}

\begin{wrapfigure}{r}{0.55\textwidth}
\centering
\includegraphics[width=0.50\textwidth]{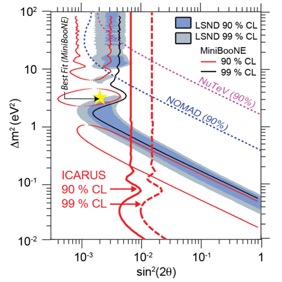}
\caption{\label{fig:icarus}ICARUS limit on sterile neutrinos.}
\end{wrapfigure}

Four decades ago, in 1977, Carlo Rubbia proposed using liquid argon TPCs as neutrino detectors.\cite{rubbia}  Research on what would become the ICARUS (Imaging Cosmic And Rare Underground Signals) started at CERN in 1985 with a 2-m$^3$ prototype and continued in 1993 in Pavia with a 10-m$^3$ prototype and finally with two 300-t modules, one of which successfully took comic data for 100 days in 2001.\cite{pavia}  The 600-t ICARUS detector was transported to Gran Sasso in 2004 and ran in the CNGS beam from 2010 to 2012.

The ICARUS collaboration had always considered the 600-t detector as just the first part of a multi-kt detector, but additional modules were not funded.  The combination of the same $L/E$ factor of over 100 that affected the OPERA experiment and the relatively small mass of the detector prevented ICARUS from doing any oscillation physics at the atmospheric mass scale, $\Delta m^2 \approx 0.0025 {\rm\ eV}^2$.  The only oscillation result from ICARUS was a limit on sterile neutrinos in the $\Delta m^2  > 0.01{\rm\ eV}^2$ range shown in Fig.~\ref{fig:icarus}.\cite{sterilefive} 

\section{Second-Generation Experiments}

\subsection{Introduction}

\begin{wrapfigure}{r}{0.40\textwidth}
\centering
\vspace{-20pt}
\includegraphics[width=0.35\textwidth]{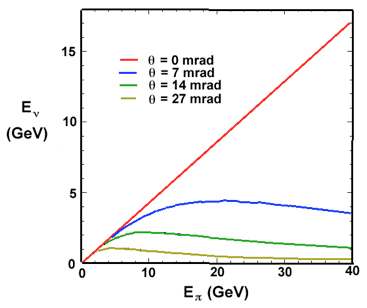}
\caption{\label{fig:offaxis}Plot of the neutrino energy versus the pion energy for several off-axis angles.}
\end{wrapfigure}

The two second-generation general experiments, T2K and \nova, are complementary experiments with similar capabilities and interests. The major difference is the baseline, 295 km for T2K and 810 km for \nova.  The difference in the baseline is significant in the determination of the mass ordering, which is measured by the effect of coherent forward scattering of \nue's on electrons in the earth.  This effect is roughly proportional to the baseline for these two experiments.

\begin{figure}[htb]
\includegraphics[width=\linewidth]{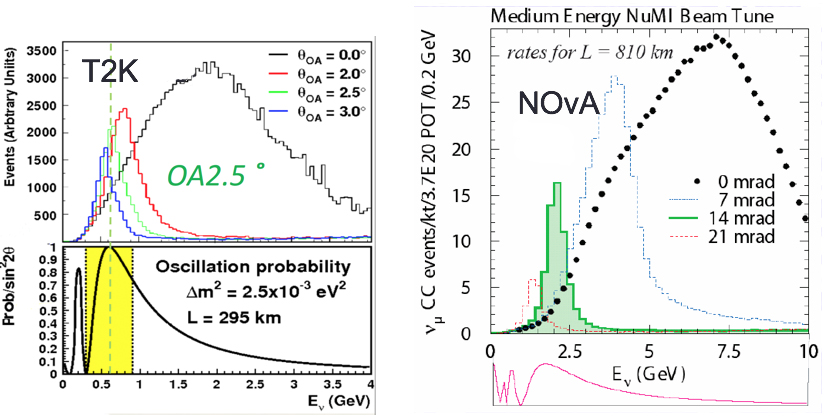}
\caption{\label{fig:beam} T2K and \nova\ beam energy  spectra for several different offaxis angles.}
\end{figure}

Both experiments place their detectors off the beam axis to create a narrow-band beam, which increase the neutrino flux near the oscillation maximum and reduces backgrounds from higher-energy neutral current events.  The reason for this is the relativistic kinematics shown in Fig.~\ref{fig:offaxis}.  Figure~\ref{fig:beam} shows the T2K\,\cite{T2Kbeam} and \nova\,\cite{novabeam} beam energy spectrum for several offaxis angles.  T2K runs at 2.5$^{\circ}$  and \nova\ runs at 14 mr off axis.

Both T2K and \nova\ use measurements of \dmsq{21}, \sinsqtwo{12}, and \sinsqtwo{13} from other experi\-ments\,\cite{pdg} in combination with their measurements of \numu\ disappearance and \nue\ appearance, in both neutrino and antineutrino beams, to gain information on the other four parameters of the standard neutrino model, $|\dmsq{32}|$ and \sinsq{23}, which are best measured in \numu\ disappearance, and $\delta_{CP}$  and the mass ordering, which are best measured in \nue\ appearance.   The lowest order equations in vacuum are
\begin{align}
& P(\numu\rightarrow\numu)\approx 1-\sinsqtwo{23}\dynamicx{32}\ldots, \\
& P(\numu\rightarrow\nue)\approx \sinsq{23}\sinsqtwo{13}\dynamicx{31}\ldots.
\end{align}

Note the difficulty in the above equations for with regard to $\theta_{23}$.  It is primarily determined by the measurement of \sinsqtwo{23} in \numu\ disappearance, but is critically used as \sinsq{23} in interpreting the measurement of \nue\ appearance.  Thus, $\theta_{23}$ is measured from a term that has a small derivative and is used in determining a double-valued term with a large derivative.

\subsection{The T2K Experiment}

Like K2K, T2K (Tokai to Kamioka) uses Super-Kamiokande as its far detector.  However, the neutrino beam is generated by the more powerful 30-GeV J-PARK proton synchrotron in Tokai, Japan.\cite{T2K}  

\begin{wrapfigure}{r}{0.60\textwidth}
\centering
\vspace{-10 pt}
\includegraphics[width=0.55\textwidth]{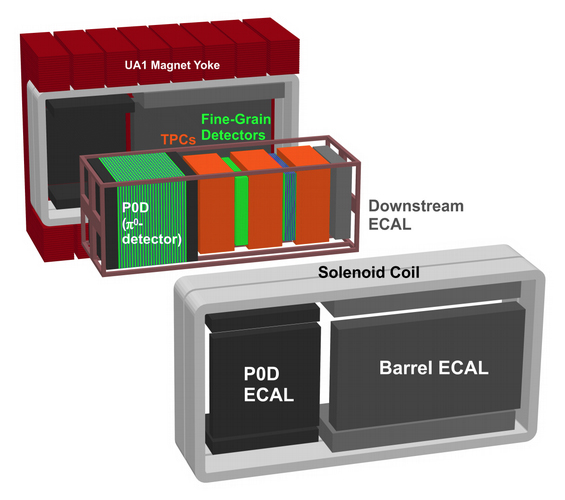}
\caption{\label{fig:T2KND}Exploaded view of the T2K near detector.}
\vspace{-20pt}
\end{wrapfigure}

The near detector at 280 m, shown in Fig.~\ref{fig:T2KND}, includes a $ \pi^0$  detector consisting of sandwiches of scintillators, water bags, and either brass or lead sheets, followed by TPCs and fine-grained scintillator bars, all included in a 0.2-T magnetic field.

T2k submitted a letter of intent in 2001.\cite{T2KLOI}  The experiment was approved in 2003 and beamline construction began in 2004.  The first run with beam came in 2010, followed by the first publication on oscillations in 2011.\cite{T2Kfirst}  Unfortunately, progress was hindered by two laboratory closures, the first due to the 2011 tsunami and the second due to an  unrelated radiation incident in 2013. \cite{incident}  

In the most recent reports, T2K has accumulated $14.9 \times 10^{20}$POT in neutrino  running and $11.2 \times 10^{20}$ POT in antineutrino  running.  The  data can be characterized by a \numu\ disappearance measurement of \dmsq{32} with a precision of $\pm 2.6\%$ and a \nue\ appearance measurement of 90 \nue CC candidates with an estimated background of 14.9 events and 9 \nuebar CC candidates with an estimated background of 4.3 events.\cite{T2Kcurrentone}  For further information and comparisons, see Table~\ref{tab:compare} and Fig.~\ref{fig:biplots}.

\subsection{The \nova\ Experiment}

The \nova\ far detector is located in northern Minnesota near the Canadian border.  It was located as far north as possible in the United States along the NUMI beamline to enhance its ability to measure the mass ordering.  

\begin{figure}[htb]
\includegraphics[width=\linewidth]{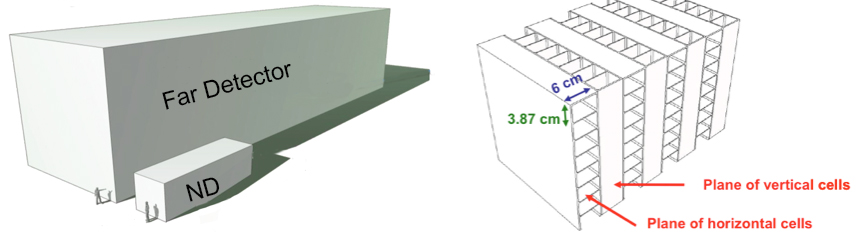}
\caption{\label{fig:novadetect} \nova\ detectors.}
\end{figure}

The 14-kt far detector consists of 344,064 $15 {\rm\ m}\times4 {\rm\ cm}\times6 {\rm\ cm}$ PVC cells filled with liquid scintillator, read out by a loop of wave-length shifting fiber with both ends connected to an avalanche photodiode.  The near detector, located about 1 km from the neutrino target is identical to the far detector, except for being much smaller.  Figure~\ref{fig:novadetect} shows the relative sizes of a human, the far detector, and the near detector and the alteration of horizontal and vertical cells.\cite{novaTDR}

Planning for the \nova\ experiment began with workshops in 2002 leading to a proposal that was submitted to the Fermilab Program Advisory Committee in March 2005\,\cite{novaproposal} and was given preliminary approval.  The proposal called for a 30 kt version of what was actually built.  At the request of the Department of Energy (DOE), the proposed \nova\ mass was downsized to 25 kt in 2006.  In June 2006, the DOE High Energy Physics Advisory Panel's P5 subpanel recommended construction of \nova\ starting in 2008.  However, in December 2007 the United States Congress cut \nova\ funding to zero. How this happened was never clear. Fermilab employees working on \nova\ were assigned to other tasks. Funding was restored in June 2008, but it took time to reassemble the team.

\nova\ construction began in 2009.  The DOE fixed the cost  at \$278M and told \nova\ to build as much far detector as possible for that amount.  The far detector pit was sized for 18 kt, but the funds ran out at 14 kt.

Since the \nova\ far detector was modular, \nova\ began taking data with 4 kt of detector in February 2014 and by November 2014 was taking data with the full 14 kt detector.  In January 2016, \nova\ nova published its first paper on oscillations.\cite{novafirst}  

\nova's most recent report on oscillation data was at the 2018 Neutrino Conference.\cite{novacurrent}   \nova\ has accumulated $8.8 \times 10^{20}$POT in neutrino running and $6.9 \times 10^{20}$POT in antineutrino  running.  The  data can be characterized by a \numu\ disappearance measurement of \dmsq{32} with a precision of $\pm 4.0\%$ and a \nue\ appearance measurement of 58 \nue CC candidates with an estimated background of 15 events and 18 \nuebar CC candidates with an estimated background of 5.3 events. 

\section{Comparisons and Other Summaries}

\subsection{Bi-event Plots}

\begin{figure}
\begin{minipage}{0.50\linewidth}
\centerline{\includegraphics[width=1.00\linewidth]{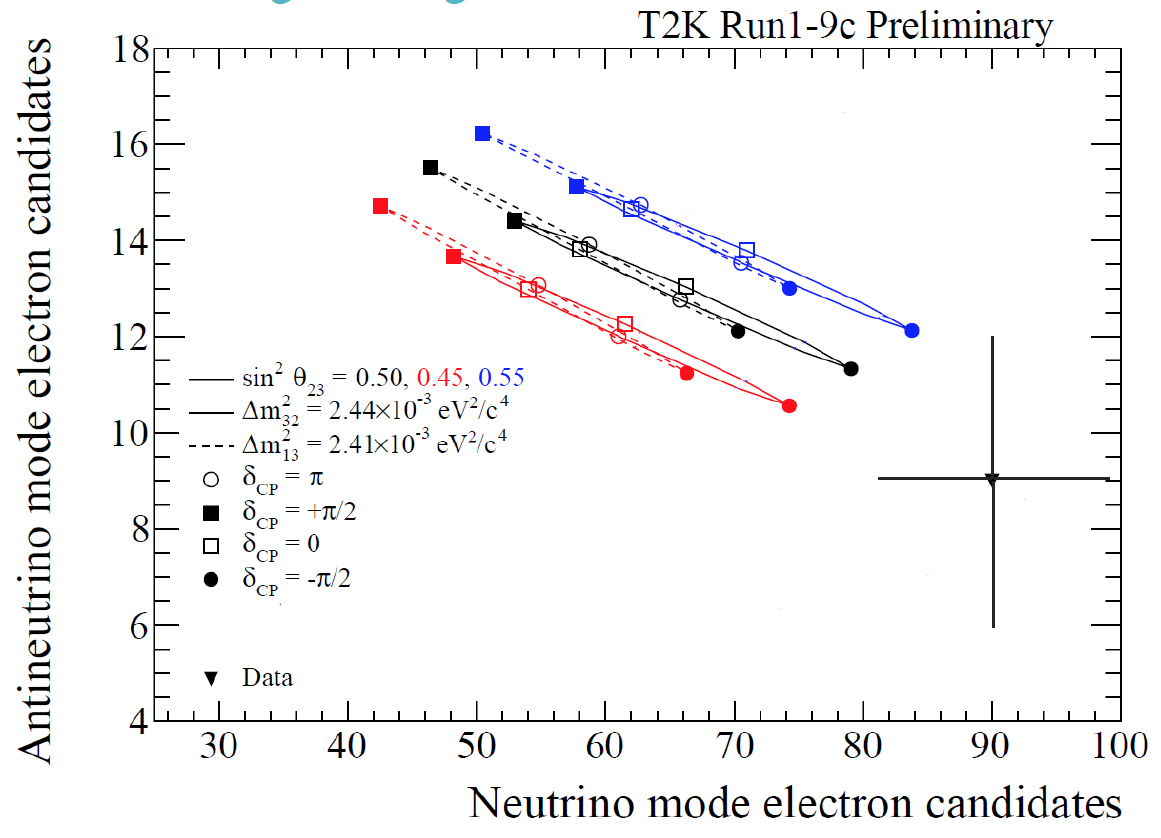}}
\end{minipage}
\hfill
\begin{minipage}{0.50\linewidth}
\centerline{\includegraphics[width=0.9\linewidth]{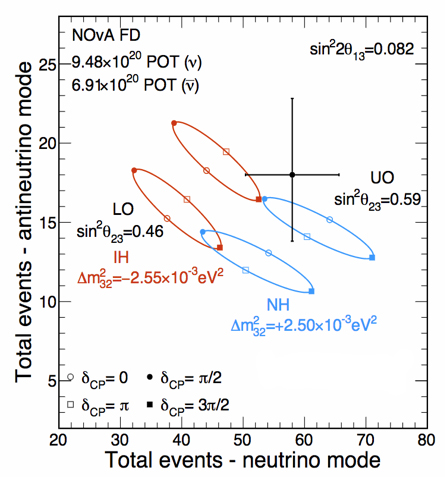}}
\end{minipage}
\hfill
\caption{Bi-event plots for T2K (left) and \nova\ (right).  In both plots, the data are shown by the point with error bars.  In the T2K plot, three sets of ellipses, representing the full range of $\delta_{CP}$,  are shown for a range of \sinsq{23} values consistent with the data.  The ellipses for the normal mass ordering are shown by solid lines and those for the inverted mass ordering are shown by dashed lines.  In the \nova\ plot, the ellipses are shown for the best fits to \sinsq{23} in the upper and lower $\theta_{23}$ octants.  The ellipses for the normal mass ordering are shown in blue and those for the inverted mass ordering are shown in red.  Note that the T2K plot from Ref.~49 has been modified to use the format of the previous T2K bi-event plot in Ref.~51 and the \nova\ plot.  See the text for additional information.}
\label{fig:biplots}
\end{figure}

As discussed above, T2K and \nova\ attempt to measure four variables simultaneously.  Even physicists working on these  experiments have difficultly visualizing how the data determines these four variables.  I believe one of the best was of aiding this visualization is the bi-event rate plot.  This plot is simply the number of \nue\ candidates on the horizontal axis versus the number of \nuebar\ candidates on the vertical axis, along with the theoretical ellipses given by $\delta_{CP}$ for each mass ordering and for characteristic values of \sinsq{23}.  These plots are shown in Fig.~\ref{fig:biplots} for T2K\,\cite{T2Kbiplot} and \nova.\cite{novacurrent}   The bi-event plot in Ref.~49 was in a confusing and not very useful form, so I took the liberty of converting it into the form that T2K had used previously\,\cite{T2Kprevious} and one that allows a direct comparison to the \nova\ plot.

I want to emphasize that these plots are pedagogical tools to help visualize where the data are located in the parameter space.  They do not represent the full analysis, which includes energy spectra and data quality.  These plots indicate that much more data is needed, and fortunately both T2K and \nova\ are likely to continue running for at least the next six years.

\subsection{Benchmarks}

Table ~\ref{tab:compare} shows a comparison of  benchmarks I have set for the quality of the \numu\ disappearance and \nue\ appearance in the four general experiments.  The benchmark for \numu\ disappearacne is the precisions of the measurement of $\Delta m^2_{32}$ and the benchmark of \nue\ appearance is a figure of merit (FoM) defined as ${\rm FoM} = s/\sqrt{s+b}$, where $s$ and $b$ are the number of signal and expected background events, respectively.

\renewcommand{\arraystretch}{1.5}
\begin{table}[h!]
\caption[]{Benchmarks for the four general experiments.  The stared value is my estimate.  See text for an explanation of the figure of merit (FoM).}
\vspace{6pt}
\label{tab:compare}
\begin{center}
\begin{tabular}{|l|c|c|c|c|}
\hline
& & & & \\
Experiment&
POT $\times 10^{20}$&
$\Delta m^2$&
~~~\nue\ FoM~~~&
Reference\\
 &
$\nu / \bar\nu$&
~~precision~~& 
\nue / \nuebar&
\\ \hline
K2K &
~0.9 / ~0.0 &
$\pm19\%^*$&
$<0 / ~-~$&
12, 14
\\ \hline
MINOS&
10.7 / ~3.4&
$\pm4.3\%$&
1.9 / 0.5&
24, 25
\\ \hline
T2K&
14.9 / 11.2&
$\pm2.6\%$&
7.9 / 2.3&
45
\\ \hline
\nova&
~8.8 / ~6.9&
$\pm4.0\%$&
5.6 / 3.0&
50
\\ \hline
\end{tabular}
\end{center}
\end{table}

\subsection{Demographics}

\begin{figure}[htb]
\includegraphics[width=\linewidth]{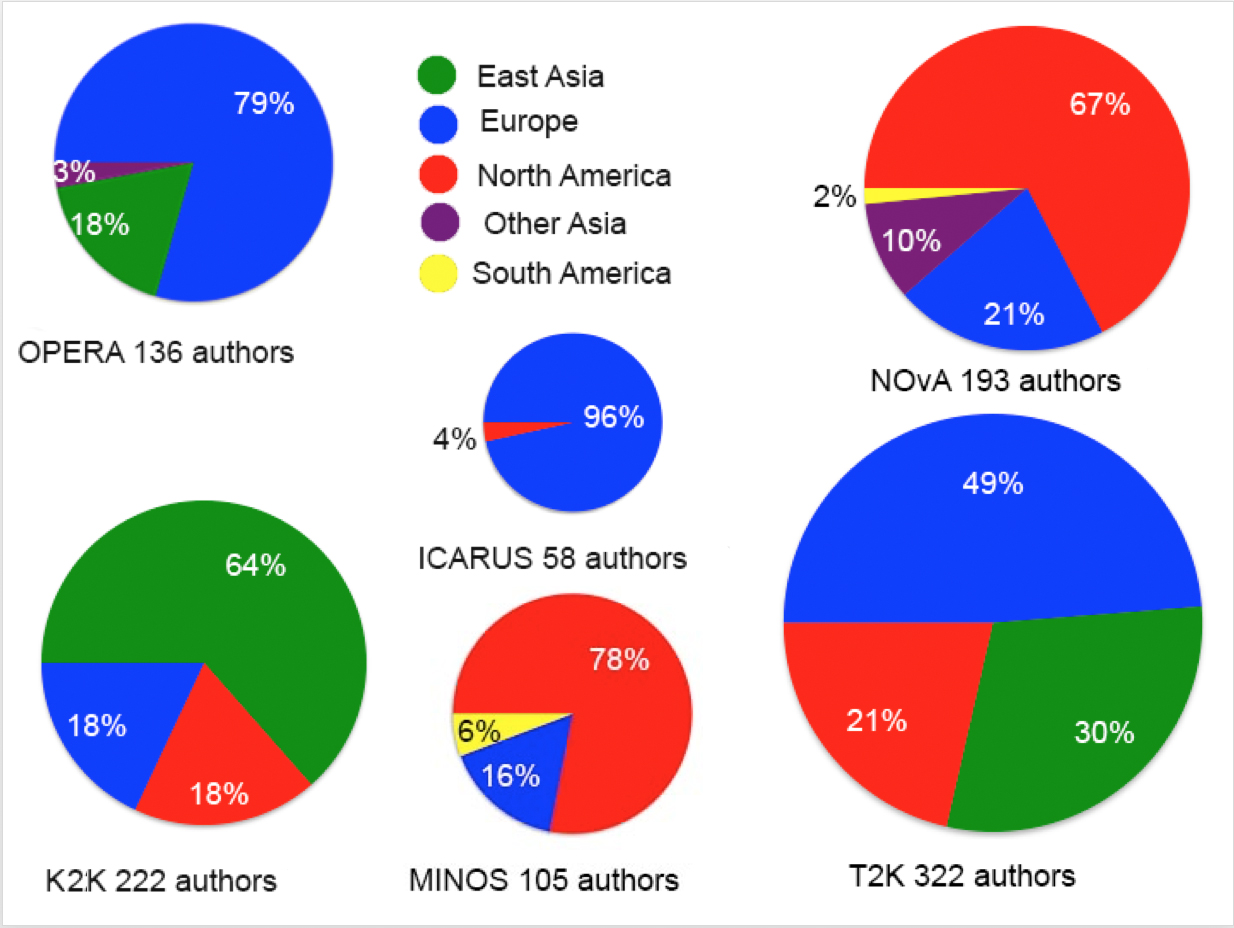}
\caption{\label{fig:demo} Demographics of the six experiment discussed in this talk as shown by their final or most current paper, whose reference are given in the text.  The areas of the pie charts are prportional to the number of authors.  ``Other Asia" refers to Israel and Turkey in the case of OPERA and to India in the case of \nova.}
\end{figure}

Figure~\ref{fig:demo} gives a summary of the demographics of each of the six collaborations by size and geographical distribution based on their final or most current paper, K2K\,\cite{k2kfinal} MINOS,\cite{MINOSfinal} OPERA,\cite{OPERA} ICARUS,\cite{sterilefive} T2K,\cite{T2Kdemo} and \nova\cite{novademo}.

\

\section*{Acknowledgments}

I thank the organizers for inviting me to review this material.  I also thank Chang Kee Jung for supplying me with Ref.~11 and for useful discussions, Maury Goodman for supplying me with Ref~17, Claudio Giganti for bringing Ref.~45 to my attention, and to Alain Blondel for useful discussions.

\end{document}